\documentclass[a4paper,fleqn,usenatbib]{mnras}

\usepackage{newtxtext,newtxmath}

\usepackage[T1]{fontenc}
\usepackage{ae,aecompl}

\usepackage{graphicx}	
\usepackage{amsmath}	
\usepackage{amssymb}	

\newcommand{\Chandra}{{\it Chandra}}
\newcommand{\Fermi}{{\it Fermi}}
\newcommand{\NICER}{{\it NICER}}
\newcommand{\RXTE}{{\it RXTE}}
\newcommand{\XMM}{{\it XMM-Newton}}
\newcommand{\psr}{PSR~J1813$-$1749}
\newcommand{\NH}{N_{\rm H}}
\newcommand{\flux}{f_{2-10}^{\rm abs}}
\newcommand{\nus}{\nu}
\newcommand{\nudot}{\dot{\nus}}
\newcommand{\nuddot}{\ddot{\nus}}
\newcommand{\Edot}{\dot{E}}
\newcommand{\Lx}{L_{\rm X}}

\title[\Chandra\ and \NICER\ observations of \psr]{Proper motion, spectra, and timing of \psr\ using \Chandra\ and \NICER}

\author[W. C. G. Ho et al.]{Wynn C. G. Ho$^{1}$\thanks{E-mail: wynnho@slac.stanford.edu},
Sebastien Guillot$^{2,3}$,
P.M. Saz Parkinson$^{4,5}$,
B. Limyansky$^{5}$,
\newauthor
C.-Y. Ng$^{6}$,
Micha{\l} Bejger$^{7}$,
Crist\'obal M. Espinoza$^{8}$,
B. Haskell$^{7}$,
\newauthor
Gaurava K. Jaisawal$^{9}$,
and C. Malacaria$^{10,11}$\thanks{NASA Postdoctoral Fellow}
\\
$^{1}$Department of Physics and Astronomy, Haverford College, 370 Lancaster Avenue, Haverford, PA, 19041, USA\\
$^{2}$IRAP, CNRS, 9 avenue du Colonel Roche, BP 44346, F-31028 Toulouse Cedex 4, France\\
$^{3}$Universit\'e de Toulouse, CNES, UPS-OMP, F-31028 Toulouse, France\\
$^{4}$Department of Physics and Laboratory for Space Research, The University of Hong Kong, Pokfulam Road, Hong Kong\\
$^{5}$Santa Cruz Institute for Particle Physics, University of California, Santa Cruz, CA, 95064, USA\\
$^{6}$Department of Physics, The University of Hong Kong, Pokfulam Road, Hong Kong\\
$^{7}$Nicolaus Copernicus Astronomical Center, Polish Academy of Sciences, ul. Bartycka 18, 00-716 Warsaw, Poland\\
$^{8}$Departamento de F\'isica, Universidad de Santiago de Chile, Avenida Ecuador 3493, 9170124 Estaci\'on Central, Santiago, Chile\\
$^{9}$National Space Institute, Technical University of Denmark, Elektrovej 327-328, DK-2800 Lyngby, Denmark\\
$^{10}$NASA Marshall Space Flight Center, NSSTC, 320 Sparkman Drive, Huntsville, AL 35805, USA\\
$^{11}$Universities Space Research Association, Science and Technology Institute, 320 Sparkman Drive, Huntsville, AL 35805, USA
}

\date{Accepted 2020 August 27. Received 2020 August 26; in original form 2020 July 2}

\pubyear{2020}

\begin{document}
\label{firstpage}
\pagerange{\pageref{firstpage}--\pageref{lastpage}}
\maketitle

\begin{abstract}
\psr\ is one of the most energetic rotation-powered pulsars known,
producing a pulsar wind nebula (PWN) and gamma-ray and TeV emission,
but whose spin period is only measurable in X-ray.
We present analysis of two \Chandra\ datasets that are separated by
more than ten years and recent \NICER\ data.
The long baseline of the \Chandra\ data allows us to derive a pulsar
proper motion
$\mu_{\rm R.A.}=-(0.\!\!\arcsec067\pm0.\!\!\arcsec010)\mbox{ yr$^{-1}$}$ and
$\mu_{\rm decl.}=-(0.\!\!\arcsec014\pm0.\!\!\arcsec007)\mbox{ yr$^{-1}$}$
and velocity $v_\perp\approx900-1600\mbox{ km s$^{-1}$}$
(assuming a distance $d=3-5$~kpc),
although we cannot exclude a contribution to the change in measured
pulsar position due to a change in brightness structure of the PWN
very near the pulsar.
We model the PWN and pulsar spectra using an absorbed power law
and obtain best-fit absorption
$\NH=(13.1\pm0.9)\times10^{22}\mbox{ cm$^{-2}$}$, photon index
$\Gamma=1.5\pm0.1$, and 0.3--10~keV luminosity
$\Lx\approx5.4\times10^{34}\mbox{ erg s$^{-1}$}(d/\mbox{ 5 kpc})^2$
for the PWN and $\Gamma=1.2\pm0.1$ and
$\Lx\approx9.3\times10^{33}\mbox{ erg s$^{-1}$}(d/\mbox{ 5 kpc})^2$ for \psr.
These values do not change between the 2006 and 2016 observations.
We use \NICER\ observations from 2019 to obtain a timing model of \psr,
with spin frequency $\nus=22.35\mbox{ Hz}$ and spin frequency
time derivative $\nudot=(-6.428\pm0.003)\times10^{-11}\mbox{ Hz s$^{-1}$}$.
We also fit $\nus$ measurements from 2009--2012 and our 2019 value
and find a long-term spin-down rate
$\nudot=(-6.3445\pm0.0004)\times10^{-11}\mbox{ Hz s$^{-1}$}$.
We speculate that the difference in spin-down rates is due
to glitch activity or emission mode switching.
\end{abstract}




\section{Introduction} \label{sec:intro}

\psr\ (also known as CXOU~J181335.16$-$174957.4) has a previously
measured spin frequency $\nus=22.37\mbox{ Hz}$ and spin-rate change
$\nudot=-6.333\times 10^{-11}\mbox{ Hz s$^{-1}$}$
\citep{gotthelfhalpern09,halpernetal12}, which make this pulsar's
spin-down energy loss rate
$\Edot=5.6\times 10^{37}\mbox{ erg s$^{-1}$}$ the fourth largest
among the 2800 known pulsars, behind only PSR~J0537$-$6910,
the Crab pulsar, and PSR~B0540$-$69 \citep{manchesteretal05}.
This large rate probably explains the pulsar's association
with the TeV source HESS~J1813$-$178 \citep{ubertinietal05}
and gamma-ray source IGR~J18135$-$1751 \citep{aharonianetal05}
and makes the pulsar an interesting target for LIGO/Virgo searches
of continuous gravitational waves \citep{abbottetal17,abbottetal19b}.
The spin frequency is only measurable at X-ray energies, as it seems to
be a variable, but unpulsed, radio source \citep{dzibetal10,dzibetal18}.
\psr\ is located in the young ($<3\mbox{ kyr}$) supernova remnant
G12.82$-$0.02 \citep{broganetal05}
at a distance $d\approx 3-5\mbox{ kpc}$ \citep{messineoetal11}.

Observations at X-ray energies are crucial to the study of \psr.
For example, the exceptional spatial resolving power of \Chandra\
allows an accurate measurement of the position of \psr\ and
separation of the X-ray spectra of the pulsar and pulsar wind
nebula (PWN) in which the pulsar is embedded.
Using a 30~ks ACIS-I observation taken in 2006,
\citet{helfandetal07} measure the pulsar position to be
R.A.$=18^{\rm h}13^{\rm m}35.\!\!^{\rm s}166$,
decl.$=-17^\circ49\arcmin57.\!\!\arcsec48$ (J2000).
They perform a spectral analysis on three spatial components, i.e.,
the PWN, an inner nebula, and the pulsar, and find each is well fit
by an absorbed power law.
More recently, \citet{townsleyetal18} analyse a number of \Chandra\
observations, including those considered here, to compile a catalog of
X-ray sources and their spectral properties, such as those of \psr.
Meanwhile,
\citet{halpernetal12} use \Chandra\ ACIS-S3 (in continuous clocking mode)
and \XMM\ EPIC-pn observations to measure the pulsar spin frequency
at three epochs, in 2009, 2011, and 2012.
While a simple linear fit of these measurements results in a
value of $\nudot$, the three observations spanning three years are
insufficient to obtain a rotation phase-connected timing model of \psr.

Here we re-analyse the 2006 \Chandra\ observation and compare it
to a set of 2016 observations.
This comparison allows us to determine not only any long-term
variability but also the pulsar proper motion over the ten-year timespan.
We present detection of the spin frequency of \psr\ using recent
\NICER\ data, which enables us to update the
timing model that is vital for the most
sensitive searches of continuous gravitational waves from this pulsar.
In Section~\ref{sec:data}, we describe our analysis procedure for
the \Chandra\ data and present results on proper motion of \psr\ and
spectral modeling of the pulsar and PWN.
In Section~\ref{sec:pulse}, we describe our pulsation search using
\NICER\ data and timing analysis of \psr.
We summarize in Section~\ref{sec:discuss}.

\section{Analysis of \Chandra\ data} \label{sec:data}

\Chandra\ observed \psr\ with ACIS-I for 30~ks on 2006 September 15
(ObsID 6685), for 13~ks on 2016 May 29 (ObsID 17695), and for 17~ks
on 2016 June 5 (ObsID 17440); see Table~\ref{tab:data}.
We reprocess data with \texttt{chandra\_repro} and Chandra Interactive
Analysis of Observations (CIAO) 4.11 and Calibration Database (CALDB)
4.8.5 \citep{fruscioneetal06}.
As in \citet{helfandetal07}, we do not account for photon pile-up
since, as they note, the maximum count rate centered on \psr\ is
$<0.004\mbox{ c s$^{-1}$}$ for ObsID 6685 and is
$<0.002\mbox{ c s$^{-1}$}$ for ObsIDs 17440 and 17695.
We note that \citet{townsleyetal18} include a pile-up correction
in their analysis of ObsID 6685 but not of 17440 and 17695.

\begin{table}
	\centering
	\caption{\Chandra\ and \NICER\ observations of \psr.}
	\label{tab:data}
	\begin{tabular}{cclc}
		\hline
		Telescope & ObsID & Date & Exposure (ks) \\
		\hline
		\Chandra & 6685 & 2006 September 15 & 30 \\
		\Chandra & 17695 & 2016 May 29 & 13 \\
		\Chandra & 17440 & 2016 June 5 & 17 \\
		\NICER & 1020440101 & 2018 August 25 & 6 \\
		\NICER & 2579030101 & 2019 June 28 & 17 \\
		\NICER & 2579030102 & 2019 June 29 & 22 \\
		\NICER & 2579030103 & 2019 June 30 & 13 \\
		\NICER & 2579030201 & 2019 July 10 & 3 \\
		\NICER & 2579030202 & 2019 July 11 & 7 \\
		\NICER & 2579030203 & 2019 July 12 & 17 \\
		\NICER & 2579030204 & 2019 July 13 & 23 \\
		\NICER & 2579030301 & 2019 July 30 & 5 \\
		\NICER & 2579030302 & 2019 July 31 & 18 \\
		\NICER & 2579030303 & 2019 August 1 & 13 \\
		\NICER & 2579030304 & 2019 August 2 & 6 \\
		\NICER & 2579030305 & 2019 August 3 & 1 \\
		\NICER & 2579030306 & 2019 August 4 & 6 \\
		\hline
	\end{tabular}
\end{table}

\subsection{Proper motion and velocity} \label{sec:motion}

We follow the recommended
procedure\footnote{https://cxc.harvard.edu/ciao/threads/reproject\_aspect/}
to improve \Chandra's astrometry.
In particular, we use \texttt{wavdetect} to detect sources in each
observation and \texttt{wcs\_match} and \texttt{wcs\_update} to match
detected sources in each 2016 observation with those detected in the
2006 observation; ObsID 6685 is used as the reference dataset given
its longer exposure time.
We then run \texttt{wavdetect} on the updated datasets of ObsID 17695
and 17440 to obtain updated pulsar positions.
The positions and 1$\sigma$ uncertainties of \psr\ are given in
Table~\ref{tab:position}.
Note that the position uncertainties determined by \texttt{wavdetect}
suggest there could be a small increase ($<15$~percent) in asymmetry in
R.A. between 2006 and 2016, perhaps due to a brightness change in the PWN,
although the PWN flux contribution near the pulsar is likely to be small.
We also note that the position in ObsID 6685 is $0.\!\!\arcsec43$ from
the position found by \citet{helfandetal07}
(see Section~\ref{sec:intro}) but is still consistent, given their
1$\sigma$ uncertainty of $0.\!\!\arcsec3$.
While we consider the absolute position (calibrated to optical
sources in the USNO-B catalog) from \citet{helfandetal07}
to be reliable, we use our measured position from ObsID 6685
for consistency in order to determine relative displacements
and extract the pulsar spectrum.

\begin{table}
	\centering
	\caption{Positions of X-ray sources. Source names are those
from \citet{townsleyetal18}, while src\# refers to the source number
in Table~1 of \citet{helfandetal07}. Number in parentheses is 1$\sigma$
error in last digit.}
	\label{tab:position}
	\begin{tabular}{rllrr}
		\hline
		ObsID & \qquad R.A. & \qquad decl. & $\Delta$R.A. & $\Delta$decl. \\
		\hline
		\multicolumn{5}{c}{CXOU~J181335.16$-$174957.4 (\psr)} \\
		 6685 & 18:13:35.151(2) & $-$17:49:57.10(3) & & \\
		17695 & 18:13:35.100(6) & $-$17:49:57.24(8) & $-0.\!\!\arcsec73(10)$ & $-0.\!\!\arcsec14(9)$ \\
		17440 & 18:13:35.112(6) & $-$17:49:57.57(7) & $-0.\!\!\arcsec55(10)$ & $-0.\!\!\arcsec47(8)$ \\
		\multicolumn{5}{c}{CXOU J181314.20$-$175343.4 (src24)} \\
		 6685 & 18:13:14.204(4) & $-$17:53:43.12(5) & & \\
		17695 & 18:13:14.200(6) & $-$17:53:43.13(12) & $-0.\!\!\arcsec05(11)$ & $-0.\!\!\arcsec01(13)$ \\
		17440 & 18:13:14.229(6) & $-$17:53:43.37(8) & $+0.\!\!\arcsec36(11)$ & $-0.\!\!\arcsec25(9)$ \\
		\multicolumn{5}{c}{CXOU J181322.48$-$175350.2 (src37)} \\
		 6685 & 18:13:22.510(13) & $-$17:53:50.00(7) & & \\
		17695 & 18:13:22.487(5) & $-$17:53:50.16(11) & $-0.\!\!\arcsec33(20)$ & $-0.\!\!\arcsec16(13)$ \\
		17440 & 18:13:22.518(5) & $-$17:53:50.13(6) & $+0.\!\!\arcsec11(20)$ & $-0.\!\!\arcsec13(9)$ \\
		\multicolumn{5}{c}{CXOU J181323.71$-$175040.5 (src41)} \\
		 6685 & 18:13:23.719(3) & $-$17:50:40.21(3) & & \\
		17695 & 18:13:23.701(6) & $-$17:50:40.41(15) & $-0.\!\!\arcsec26(10)$ & $-0.\!\!\arcsec20(15)$ \\
		17440 & 18:13:23.722(9) & $-$17:50:40.36(13) & $+0.\!\!\arcsec04(13)$ & $-0.\!\!\arcsec15(13)$ \\
		\multicolumn{5}{c}{CXOU J181341.20$-$175115.4 (src65)} \\
		 6685 & 18:13:41.210(8) & $-$17:51:15.11(17) & & \\
		17695 & 18:13:41.158(16) & $-$17:51:15.66(19) & $-0.\!\!\arcsec74(25)$ & $-0.\!\!\arcsec55(25)$ \\
		17440 & 18:13:41.190(26) & $-$17:51:15.77(30) & $-0.\!\!\arcsec28(38)$ & $-0.\!\!\arcsec66(34)$ \\
		\hline
	\end{tabular}
\end{table}

\begin{figure*}
\begin{center}
\includegraphics[width=1.6\columnwidth]{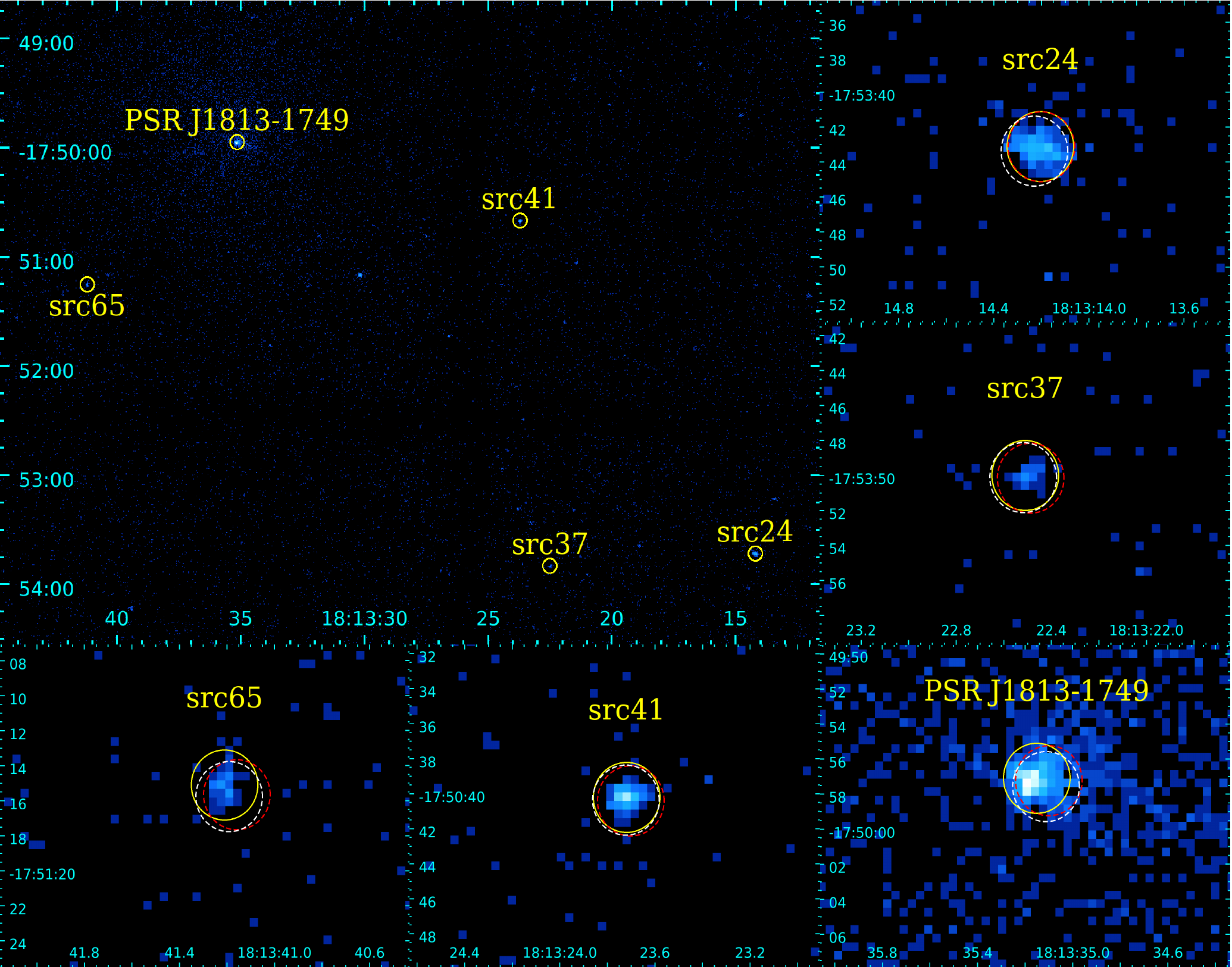}
\caption{
\Chandra\ 2006 image (ObsID 6685) of \psr\ and four nearby and bright
sources that are used to refine the astrometry between 2006 and 2016.
The latter four sources are labeled by their source number from Table~1 of
\citet{helfandetal07} (see also Table~\ref{tab:position}).
Solid yellow, dashed red, and dashed white circles
(with radius of 2\arcsec\ in all zoomed-in panels)
indicate positions of each source in 2006, 2016 (ObsID 17695), and 2016
(ObsID 17440), respectively.
}
\label{fig:motion}
\end{center}
\end{figure*}

From Table~\ref{tab:position}, we see that the position of \psr\
after astrometric correction appears to be displaced to the southwest
by $\approx 0.\!\!\arcsec7$ between 2006 and 2016.
To refine this displacement, we select four of the nearest
(to the pulsar) and brightest sources and use their position
and 1$\sigma$ uncertainty in each dataset as determined by
\texttt{wavdetect}.
These sources are shown in Figure~\ref{fig:motion} and their positions
are given in Table~\ref{tab:position};
source names and labels are from \citet{helfandetal07,townsleyetal18}.
There are a few other sources that are brighter
(in ObsID 6685) than some of the four, but these other sources
are fainter than the four chosen here in ObsID 17695 and/or 17440.
We calculate a weighted (by square of uncertainty) mean shift of
these four sources: ($\Delta$R.A.,$\Delta$decl.)
$=(-0.\!\!\arcsec22\pm0.\!\!\arcsec07,-0.\!\!\arcsec16\pm0.\!\!\arcsec07)$
for ObsID 17695 and
$(+0.\!\!\arcsec19\pm0.\!\!\arcsec08,-0.\!\!\arcsec20\pm0.\!\!\arcsec06)$
for ObsID 17440.
Accounting for this shift, we obtain a weighted mean displacement
($\Delta$R.A.,$\Delta$decl.)
$=(-0.\!\!\arcsec62\pm0.\!\!\arcsec09,-0.\!\!\arcsec15\pm0.\!\!\arcsec08)$
for \psr.
Considering the 9.7~year time difference between 2006 and 2016 observations,
we determine that \psr\ appears to be moving with a proper motion of
$\mu_{\rm R.A.}=-(0.\!\!\arcsec067\pm0.\!\!\arcsec010)\mbox{ yr$^{-1}$}$
and
$\mu_{\rm decl.}=-(0.\!\!\arcsec014\pm0.\!\!\arcsec007)\mbox{ yr$^{-1}$}$,
after accounting for $\cos$(decl.) in the apparent R.A. motion of
$-(0.\!\!\arcsec064\pm0.\!\!\arcsec009)\mbox{ yr$^{-1}$}$.
For an uncertain distance of $\approx 3-5\mbox{ kpc}$
\citep{messineoetal11,halpernetal12},
the proper motion implies a transverse velocity
$v_\perp\approx 900-1600\mbox{ km s$^{-1}$}$, with an uncertainty
of $\sim 300\mbox{ km s$^{-1}$}$.
This velocity is high but not extraordinary compared to that measured
for other neutron stars \citep{kargaltsevetal17,delleretal19,dangetal20}.
Finally, the pulsar is $\sim 20\arcsec$ from the center of
the supernova remnant G12.82$-$0.02 (see, e.g., \citealt{dzibetal18}),
and therefore this velocity would indicate a pulsar age of
$\sim 300\mbox{ yr}$, which is at the lower end of the age range of
200--3000~yr for the remnant \citep{broganetal05}.

\subsection{Spectra} \label{sec:spectra}

\begin{figure}
\begin{center}
\includegraphics[width=\columnwidth]{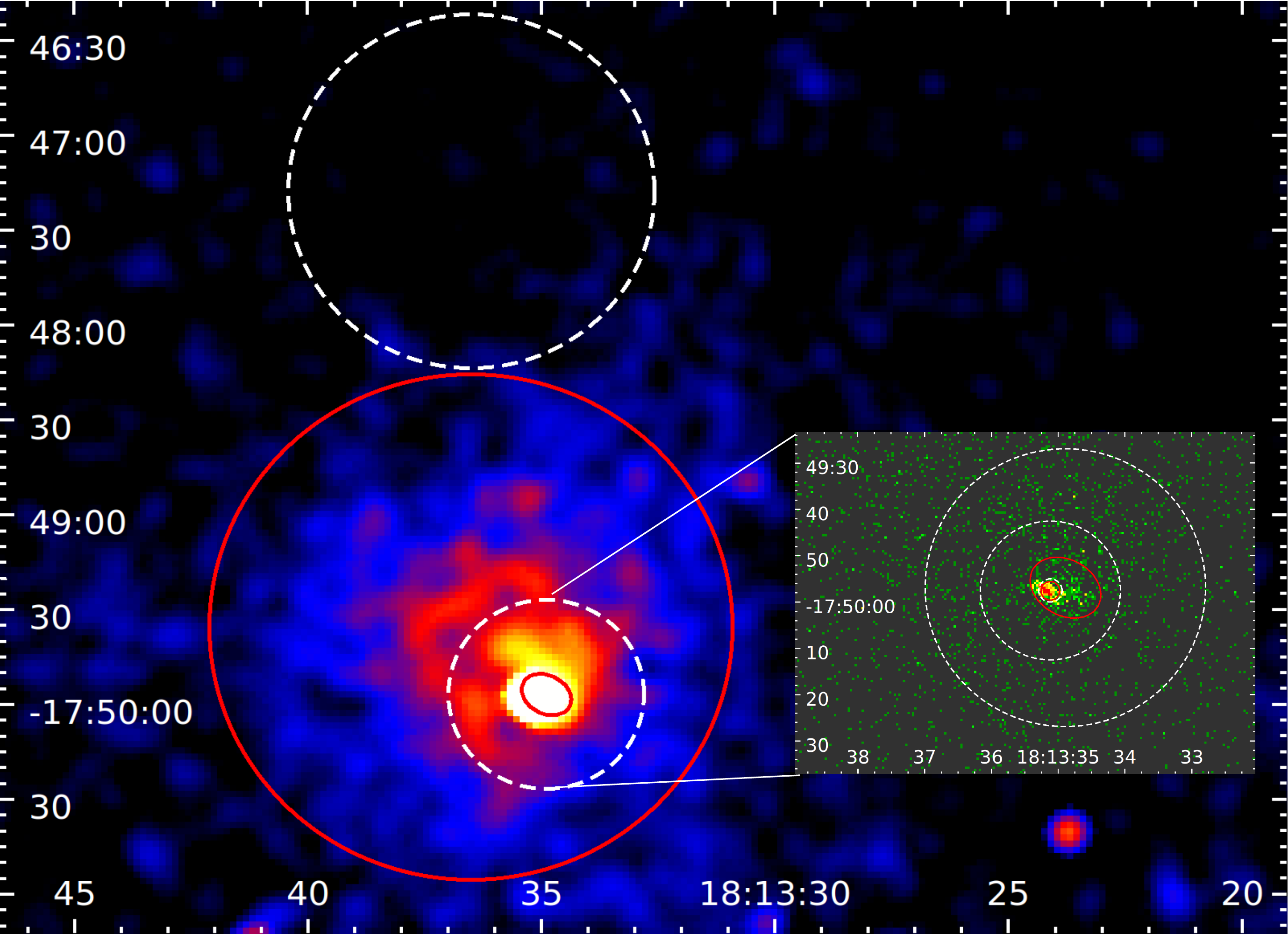}
\caption{
\Chandra\ image (ObsID 17440) of the PWN, inner nebula, and \psr;
0.5--7~keV image is smoothed to make the PWN more visible.
Inset: Zoomed-in unsmoothed view of inner nebula and \psr.
Solid curves indicate extraction regions for the PWN (80\arcsec\ radius),
inner nebula ($6\arcsec\times8\arcsec$ radii ellipse),
and pulsar (2\arcsec\ radius) spectra.
Dotted curves indicate background extraction regions for the
PWN (56\arcsec\ radius), inner nebula (30\arcsec\ radius), and pulsar
(2.\!\!\arcsec5--15\arcsec\ radii annulus).
Embedded regions are excluded from each source's spectral extraction
(see text for details).
}
\label{fig:fov}
\end{center}
\end{figure}

Spectra are extracted using \texttt{specextract} from regions shown
in Figure~\ref{fig:fov}
(from ObsID 17440; see also Figure~13 of \citealt{kuiperhermsen15}
for an image from ObsID 6685) for the PWN, inner nebula, and pulsar.
These regions are chosen to be the same as those used by \citet{helfandetal07}
and are different from that used in \citet{townsleyetal18} for the pulsar.
Specifically, source counts for the PWN are from a 80\arcsec\ radius circle
centered on R.A.$=18^{\rm h}13^{\rm m}36.\!\!^{\rm s}50$,
decl.$=-17^\circ49\arcmin35.\!\!\arcsec6$
and excluding the inner nebula source region,
while background for the PWN is from a 56\arcsec\ radius circle centered on
R.A.$=18^{\rm h}13^{\rm m}36.\!\!^{\rm s}50$,
decl.$=-17^\circ47\arcmin17.\!\!\arcsec6$,
i.e., $2.\!\!\arcmin3$ north of source region.
Source counts for the inner nebula are from a $6\arcsec\times8\arcsec$
radii ellipse centered on R.A.$=18^{\rm h}13^{\rm m}34.\!\!^{\rm s}89$,
decl.$=-17^\circ49\arcmin56.\!\!\arcsec9$
and excluding the pulsar source region,
while background for the inner nebula is from a 30\arcsec\ radius circle
centered on it and excluding the inner nebula region.
Source counts for the pulsar are from a 2\arcsec\ radius circle
(which encircles an energy fraction of 90~percent at 4.5~keV)
centered on its position as given in Table~\ref{tab:position},
while background for the pulsar is from a $2.\!\!\arcsec5-15\arcsec$
radius circular annulus around the pulsar.
Since ObsIDs 17440 and 17695 are taken only one week apart, we merge
spectra extracted from these two observations using
\texttt{combine\_spectra} and \texttt{dmgroup}.
PWN, inner nebula, and pulsar spectra are binned with a minimum
of 30, 15, and 20 counts per energy bin, respectively.
Fit results for the PWN and pulsar are the same within uncertainties
when using a minimum of 15 counts per bin, as done in \citet{helfandetal07}.

We perform spectral fitting using Xspec 12.10.1 \citep{arnaud96}.
We use an absorbed power law (PL) model composed of \texttt{tbabs}
and \texttt{powerlaw}.
The former is to model photoelectric absorption in the interstellar
medium, with abundances from \citet{wilmsetal00} and cross-sections
from \citet{verneretal96};
note that use of \texttt{phabs}, instead of \texttt{tbabs}, leads to
very similar results except for a slightly higher best-fit $\NH$
($=13.4\times 10^{22}\mbox{ cm$^{-2}$}$ for the PWN).
The power law is to model the intrinsic spectrum of the pulsar or
pulsar wind.
For each set of spectra (PWN, inner nebula, and pulsar), we conduct
two different fits.
For the PWN, we first allow varying but linked values of absorption
and photon index between the 2006 and 2016 data, so that only the
power law normalization varies between these two epochs.
For the inner nebula and pulsar, our first fit fixes the absorption
to the best-fit value of the PWN spectral model,
i.e., $\NH=13.1\times 10^{22}\mbox{ cm$^{-2}$}$
(see Table~\ref{tab:spectra}), but allows a varying but linked value
of the photon index.
The second fits have all model parameters free to vary and untied
between observations.

\begin{table}
	\centering
\caption{Spectral fits with absorbed power law.
Absorption $\NH$ is in $10^{22}\mbox{ cm$^{-2}$}$, power law
normalization is in $10^{-4}\mbox{ photon cm$^{-2}$ s$^{-1}$ keV$^{-1}$}$,
and absorbed 2--10~keV flux
$\flux$ is in $10^{-12}\mbox{ erg cm$^{-2}$ s$^{-1}$}$.
Errors are 1$\sigma$, and parameter values without errors are fixed.}
	\label{tab:spectra}
	\begin{tabular}{cccccc}
		\hline
		Year & $\NH$ & $\Gamma$ & PL norm. & $\flux$ & $\chi^2$/dof \\
		\hline
		\multicolumn{6}{c}{PWN} \\
		2006 & $13.11_{-0.86}^{+0.89}$ & $1.46\pm{0.12}$ & $20.3_{-3.9}^{+4.9}$ & $7.5_{-0.3}^{+0.1}$ & 330/308 \\
		2016 & & & $20.7_{-3.9}^{+5.0}$ & $7.7_{-0.3}^{+0.1}$ & \\
		\\
		2006 & $12.6_{-1.2}^{+1.3}$ & $1.32_{-0.16}^{+0.17}$ & $16.0_{-4.0}^{+5.6}$ & $7.8_{-0.6}^{+0.1}$ & 173/160 \\
		2016 & $13.8_{-1.2}^{+1.3}$ & $1.63_{-0.17}^{+0.18}$ & $27.1_{-7.1}^{+10.0}$ & $7.4_{-0.9}^{+0.1}$ & 153/146 \\
		\\ \multicolumn{6}{c}{inner nebula} \\
		2006 & $13.1$ & $0.77\pm{0.17}$ & $0.40_{-0.10}^{+0.12}$ & $0.54_{-0.07}^{+0.04}$ & 56.0/54 \\
		2016 & & & $0.93_{-0.22}^{+0.29}$ & $1.2\pm{0.1}$ & \\
		\\
		2006 & $14.9_{-8.1}^{+9.2}$ & $0.47_{-0.76}^{+0.84}$ & $0.27_{-0.27}^{+1.1}$ & $\lesssim0.6$ & 19.4/18 \\
		2016 & $13.8_{-4.4}^{+4.8}$ & $1.03_{-0.46}^{+0.49}$ & $1.4_{-0.8}^{+2.1}$ & $1.2_{-1.0}^{+0.1}$ & 33.5/33 \\
		\\ \multicolumn{6}{c}{\psr} \\
		2006 & $13.1$ & $1.24\pm{0.11}$ & $2.79_{-0.42}^{+0.49}$ & $1.6\pm{0.1}$ & 55.8/66 \\
		2016 & & & $2.77_{-0.43}^{+0.50}$ & $1.5\pm{0.1}$ & \\
		\\
		2006 & $16.5_{-3.6}^{+3.9}$ & $1.44_{-0.42}^{+0.44}$ & $4.3_{-2.3}^{+5.4}$ & $1.6_{-0.7}^{+0.1}$ & 34.7/37 \\
		2016 & $14.7_{-3.8}^{+4.1}$ & $1.67_{-0.46}^{+0.50}$ & $5.6_{-3.6}^{+9.2}$ & $1.4_{-0.9}^{+0.1}$ & 16.1/26 \\
		\hline
	\end{tabular}
\end{table}

The results of our different spectral fits are given in
Table~\ref{tab:spectra}.
A comparison between the fit where all PWN parameters are free to vary
and the fit where some PWN parameters are tied between observations
yields a $F$-test probability of 22~percent for the fit improvement of the
former to be produced by chance.
$F$-test comparisons for fits to the inner nebula and pulsar spectra
yield probabilities of 41~percent and 11~percent, respectively.
These probabilities indicate that tied values of $\NH$ and photon
index $\Gamma$ are sufficient to describe the combined spectra, and
we also see that all model parameter values for both sets of fits
are within each other's uncertainties.
Thus $\NH$ and $\Gamma$ do not change between 2006 and 2016 observations.
While the PWN and pulsar fluxes are also the same, the inner
nebula flux may be different, although uncertainties in the precise
spectral extraction and inferred values allow for consistency
between 2006 and 2016.

\begin{figure}
\begin{center}
\includegraphics[width=0.7\columnwidth,angle=-90,trim=30 40 0 0,clip]{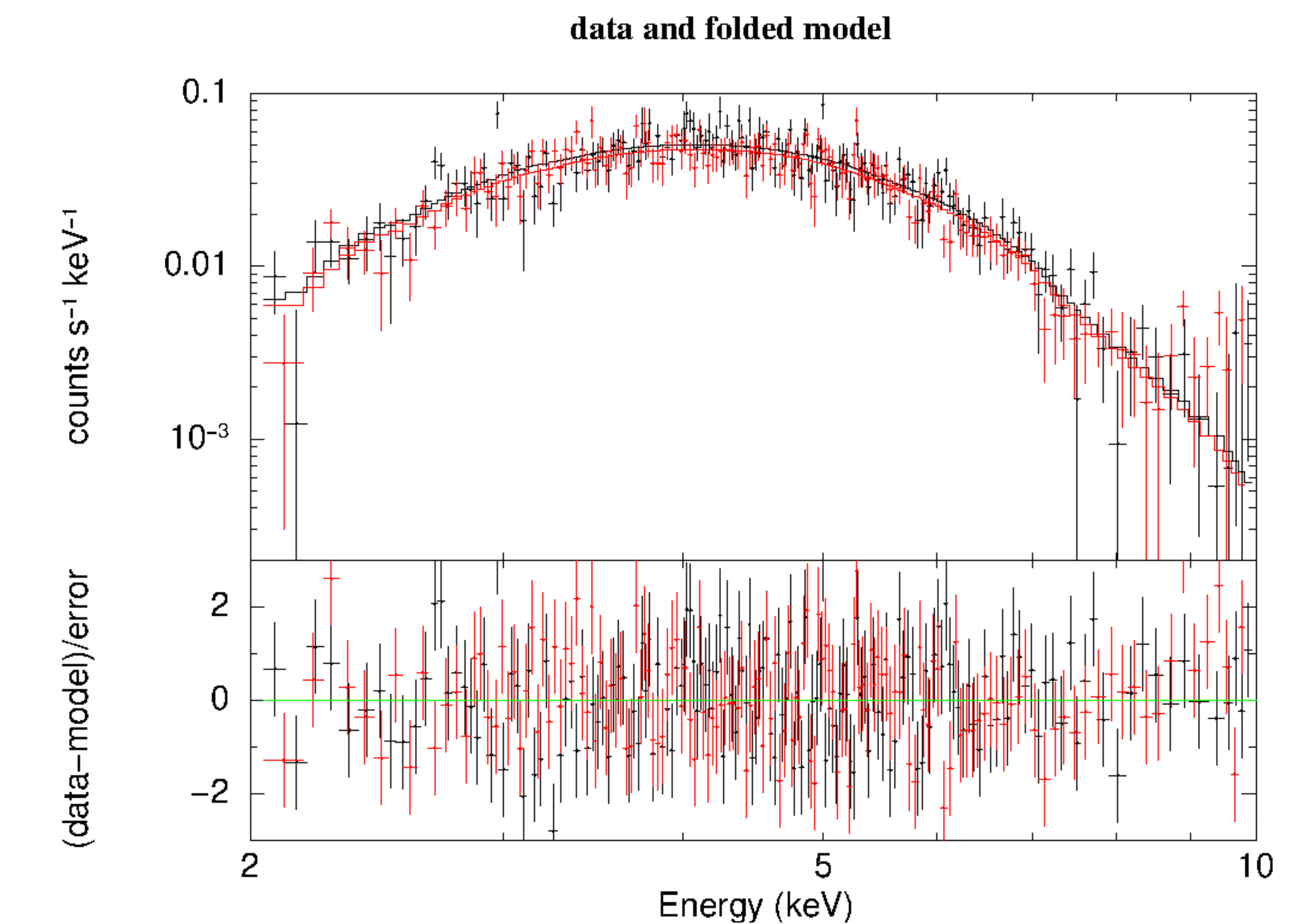}
\includegraphics[width=0.7\columnwidth,angle=-90,trim=30 40 0 0,clip]{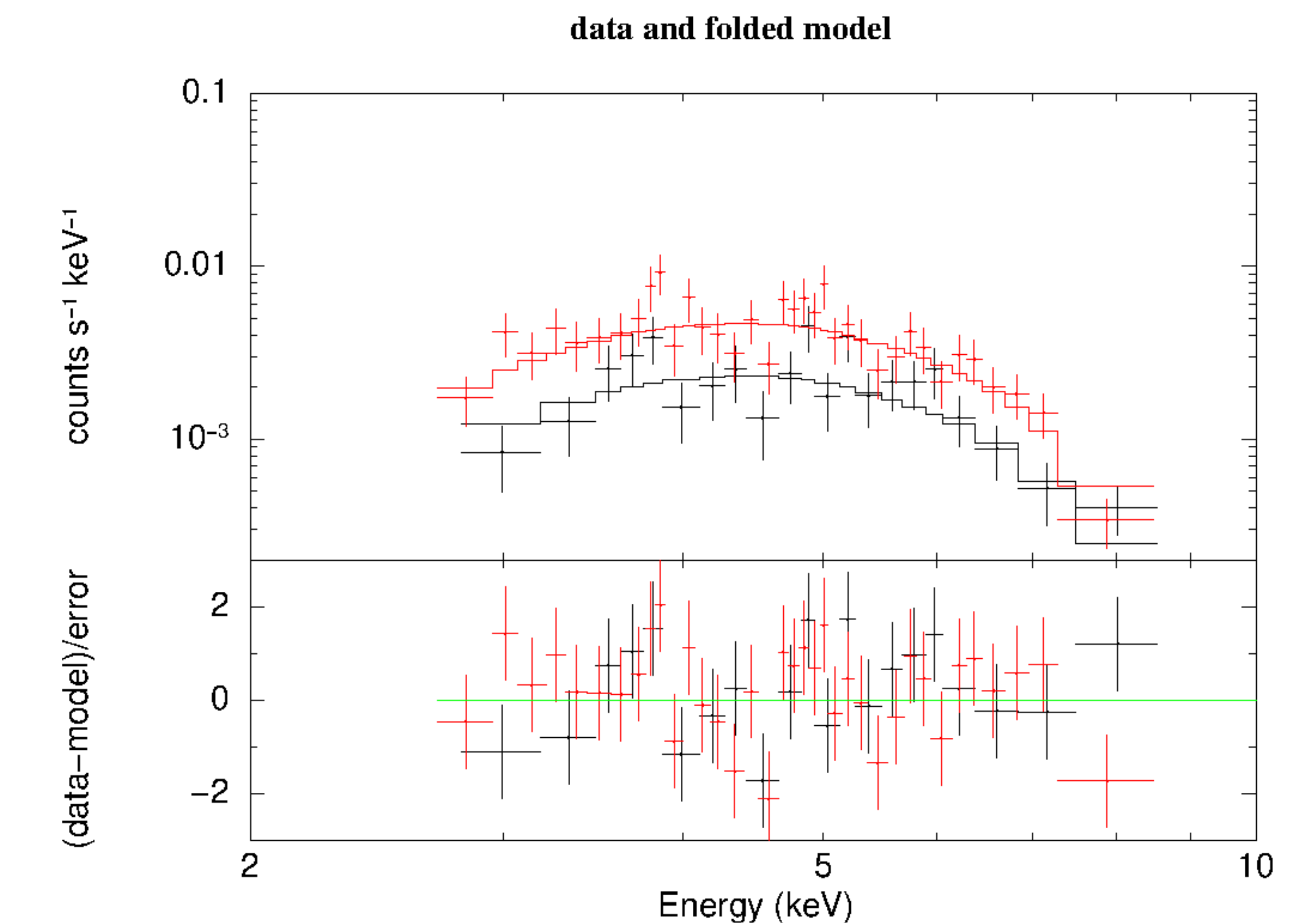}
\includegraphics[width=0.7\columnwidth,angle=-90,trim=30 40 0 0,clip]{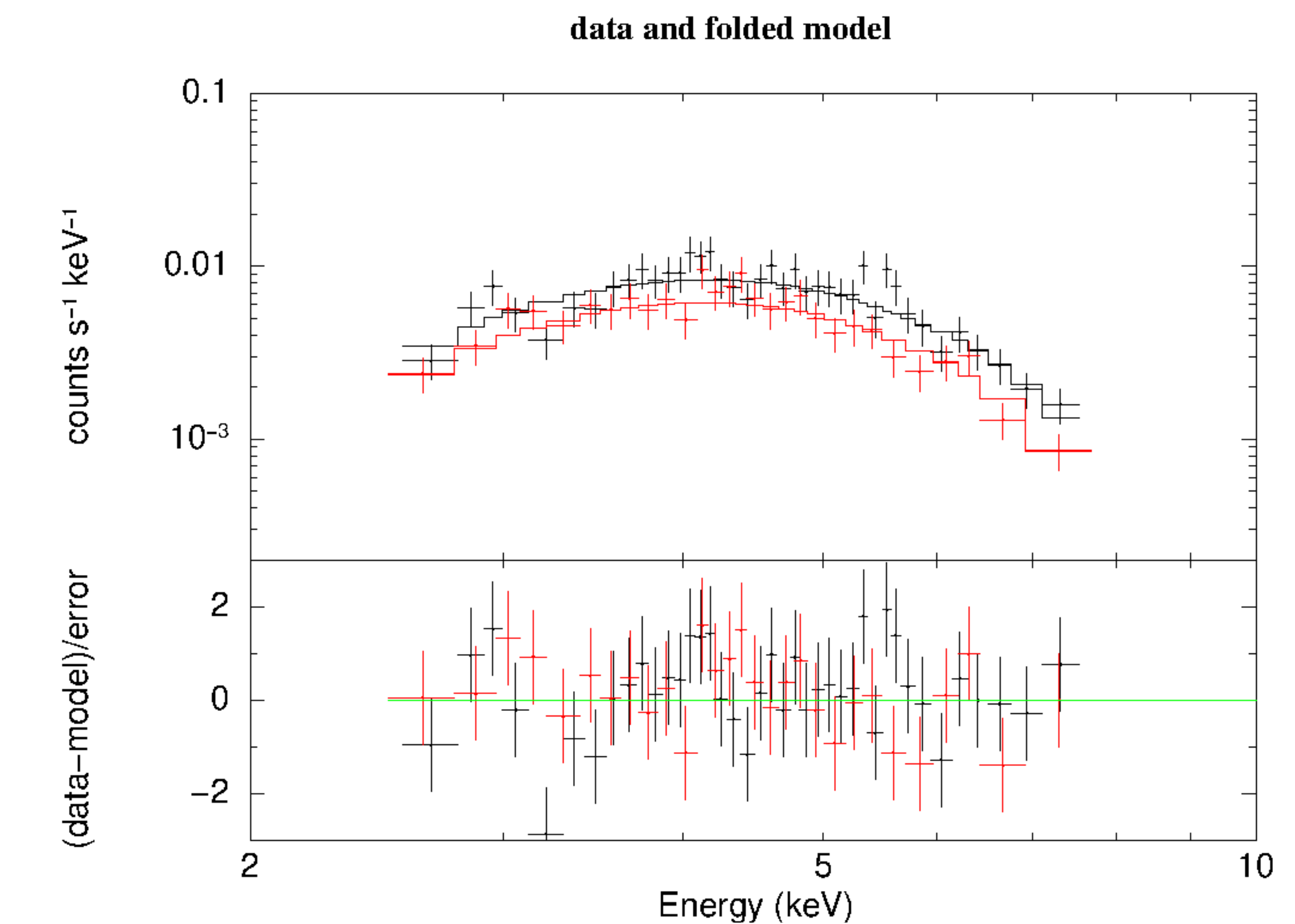}
\caption{
Upper panels show \Chandra\ spectra (black for ObsID 6685 and red
for merged 17695/17440) of the PWN (top), inner nebula (middle),
and \psr\ (bottom) and best-fit absorbed power law model.
The model is from the analysis with $\NH$ and $\Gamma$ tied between
observations and $\NH$ for the inner nebula and pulsar fixed to the
best-fit value ($\NH=13.1\times 10^{22}\mbox{ cm$^{-2}$}$) for the PWN.
Lower panels show fit residuals.
}
\label{fig:spectra}
\end{center}
\end{figure}

Figure~\ref{fig:spectra} shows the PWN, inner nebula, and pulsar
spectra and best-fit models, with model parameters tied between
observations.
For the PWN, $\NH=(13.1\pm0.9)\times10^{22}\mbox{ cm$^{-2}$}$,
$\Gamma=1.5\pm0.1$, and absorbed 2--10~keV flux
$\flux\approx7.6\times10^{-12}\mbox{ erg cm$^{-2}$ s$^{-1}$}$.
The flux is constant across the 10 years between observations and
equates to an unabsorbed 0.3--10~keV flux
$f_{\rm X}=1.8\times10^{-11}\mbox{ erg cm$^{-2}$ s$^{-1}$}$ and
luminosity $\Lx=5.4\times10^{34}\mbox{ erg s$^{-1}$}\,(d/\mbox{ 5 kpc})^2$.
For \psr, $\Gamma=1.2\pm0.1$ and
$\flux\approx1.5\times10^{-12}\mbox{ erg cm$^{-2}$ s$^{-1}$}$,
which results in
$f_{\rm X}=3.1\times10^{-12}\mbox{ erg cm$^{-2}$ s$^{-1}$}$ and
$\Lx=9.3\times10^{33}\mbox{ erg s$^{-1}$}\,(d/\mbox{ 5 kpc})^2$.
The X-ray luminosity relative to spin-down power of \psr\ is
$\Lx/\Edot=1.7\times 10^{-4}$ (assuming a distance of 5~kpc),
and it is $\Lx\mbox{(PWN)}/\Edot=9.6\times 10^{-4}$ for the PWN;
these are typical for rotation-powered pulsars
\citep{beckertrumper97,becker09,enotoetal19}.

We briefly compare our spectral fitting results to those of previous works.
\citet{helfandetal07} fit the PWN spectra (ObsID 6685) with
absorption $\NH=9.8_{-0.9}^{+1.2}\times 10^{22}\mbox{ cm$^{-2}$}$,
$\Gamma=1.3\pm0.3$ (errors are at 90~percent confidence), and
$\flux=5.6\times 10^{-12}\mbox{ erg cm$^{-2}$ s$^{-1}$}$.
Spectra of the inner nebula and pulsar are each fit with a PL and
fixing $\NH$ to that of the PWN.
The inner nebula spectral fit yields $\Gamma=0.4_{-0.7}^{+0.4}$ and
$\flux=4\times 10^{-13}\mbox{ erg cm$^{-2}$ s$^{-1}$}$,
while the pulsar spectral fit yields $\Gamma=1.3\pm0.3$ and
$\flux=1.3\times 10^{-12}\mbox{ erg cm$^{-2}$ s$^{-1}$}$.
\citet{kuiperhermsen15} analyse a 98~ks \XMM\ EPIC-pn spectrum taken
on 2009 March 27 (ObsID 0552790101) and,
with an extraction radius of 15\arcsec\ which includes some contribution
from the PWN, find a consistent photon index $\Gamma=1.31\pm0.01$ and
flux $\flux=(1.94\pm0.11)\times10^{-12}\mbox{ erg cm$^{-2}$ s$^{-1}$}$
but somewhat higher absorption 
$\NH=(11.7\pm0.35)\times 10^{22}\mbox{ cm$^{-2}$}$.
\citet{townsleyetal18} fit the 2006 and 2016 \Chandra\ spectral data
used here and find $\NH=(17\pm2)\times 10^{22}\mbox{ cm$^{-2}$}$,
$\Gamma=1.6\pm0.3$, and absorbed 2--8~keV flux
$f^{\rm abs}_{2-8}=(0.8-0.9)\times 10^{-12}\mbox{ erg cm$^{-2}$ s$^{-1}$}$.

Other than absorption $\NH$, our best-fit model parameter values for
the PWN, inner nebula, and pulsar agree with those of \citet{helfandetal07}
in their analysis of just the 2006 \Chandra\ data, except for a
$\approx 20$~percent difference in inferred flux of the PWN,
and with the parameter values of \citet{kuiperhermsen15} and
\citet{townsleyetal18}.
For $\NH$, our fit to the PWN spectrum using either the 2006, 2016,
or combined data yields
$\NH\sim(12-15)\times10^{22}\mbox{ cm$^{-2}$}$,
which is significantly higher than \citet{helfandetal07} find from
their fit to the 2006 observation, i.e.,
$\approx(9-11)\times10^{22}\mbox{ cm$^{-2}$}$
[see also \citealt{marellietal11}, who find
$\approx(5-13)\times10^{22}\mbox{ cm$^{-2}$}$].
Note that the Galactic H{\sc i} column density in the direction of \psr\
implies only $\NH=1.7\times10^{22}\mbox{ cm$^{-2}$}$ \citep{hi4pi16}.
A similarly high absorption, albeit with large uncertainties, is
obtained in our fit to the pulsar (or inner nebula) 2006 or 2016
spectrum, which is in agreement with
$\sim17\times10^{22}\mbox{ cm$^{-2}$}$ from the pulsar spectral
fits of \citet{townsleyetal18}.
In addition, fits of \XMM\ data, which cannot fully separate PWN
and pulsar emission components, yield a high absorption of
$\NH\approx(10-13)\times10^{22}\mbox{ cm$^{-2}$}$ \citep{funketal07}
and $(11-12)\times10^{22}\mbox{ cm$^{-2}$}$ \citep{kuiperhermsen15}.
Therefore, a higher value of $\NH$ than that found by \citet{helfandetal07}
is likely more indicative of the X-ray absorption of the PWN and pulsar
(see also discussion of $\NH$ and its implication on the distance to
\psr\ in \citealt{halpernetal12}).

\section{Timing analysis of \NICER\ data} \label{sec:pulse}

We process and filter \NICER\ data of \psr\ (see Table~\ref{tab:data})
using HEASoft~6.26.1, NICERDAS~2019-06-19\_V006a,
and the \texttt{psrpipe.py} script from the NICERsoft package\footnote{\url{https://github.com/paulray/NICERsoft}\label{foot:nicersoft}}.
We exclude all events from ``hot'' detector 34, which gives elevated
count rates in some circumstances, and portions of exposure accumulated
during passages through the South Atlantic Anomaly.
\NICER\ experienced a time stamp anomaly which resulted in incorrect
time stamps for data taken with MPU1 between 2019 July 8 and 23;
we follow the recommended procedure for excluding MPU1 data
(for only ObsIDs 2579030201--4) from our
analysis\footnote{\url{https://heasarc.gsfc.nasa.gov/docs/nicer/data_analysis/nicer_analysis_tips.html\#July2019-MPU1_Timing_Errors}}.
Using these filtering criteria, we obtain a total of 301,084 events
for pulse timing analysis with individual exposure times shown in
Table~\ref{tab:data} and a total exposure time of 159~ks.
We run \texttt{barycorr} to transform between Terrestrial Time, used
for event time stamps, and Barycentric Dynamical Time (TDB).  We adopt
the JPL-DE405 solar system ephemeris and absolute sky position measured
using the most recent \Chandra\ observation (ObsID~17440 from 2016 June 5),
i.e., R.A.$=18^{\rm h}13^{\rm m}35.\!\!^{\rm s}112$,
decl.$=-17^\circ49\arcmin57.\!\!\arcsec57$ (J2000).

\NICER\ is sensitive to 0.25--12~keV photons.
Previous measurements of the 44.7~ms spin period of \psr\ are made
at 2--10~keV using \Chandra\ and \XMM\ with a pulsed fraction
$\approx 50$~percent \citep{gotthelfhalpern09,halpernetal12,kuiperhermsen15}
and at $\sim$2--27~keV using \RXTE, with pulsations being stronger at
lower energies \citep{kuiperhermsen15}.
As we show in Section~\ref{sec:spectra} and in agreement with
previous works, the pulsar spectrum suffers from strong
interstellar absorption ($\NH\sim 10^{23}\mbox{ cm$^{-2}$}$).
Therefore we initially select data in the 1--10~keV range, resulting
in 218,624 events.
We do not conduct spectral analyses using \NICER\ data since the
large non-imaging field of view implies an extracted spectrum will
primarily be due to that of the PWN and supernova remnant, and
spectra from \Chandra\ and \XMM\ of these extended sources are
presented in other studies \citep{funketal07,helfandetal07}.

We perform a blind pulsation search on the merged dataset
of all 14 \NICER\ observations, using the time differencing technique
applied to gamma-ray data in previous works
\citep{atwoodetal06,abdoetal09,sazparkinsonetal10}.
We use a time window of 524,288 seconds
(Fast Fourier Transform size $=67108864$, with resolution of
$1.90735\times10^{-6}\mbox{ Hz}$) and scan $\nudot/\nus$ between
0 and $1.300\times10^{-11}\mbox{ Hz}$ in 6494 steps of
$2.002\times10^{-15}\mbox{ Hz}$.
The best pulsation candidate has a frequency
$\nus=22.35194397\mbox{ Hz}$ and
$\nudot=-6.480\times10^{-11}\mbox{ Hz s$^{-1}$}$
at MJD~58527.51318287 (mid-point of observations),
with a p-value of $8.9\times10^{-7}$.
Recall that \citet{halpernetal12} determine an incoherent timing
model with $\nus=22.3717124\mbox{ Hz}$ and 
$\nudot=-6.333\times10^{-11}\mbox{ Hz s$^{-1}$}$ at MJD~54918.14;
this $\nudot$ leads to a frequency change $\Delta\nu=-0.01975$~Hz
by the time of the \NICER\ observations and an expected
$\nus=22.35196\mbox{ Hz}$, which closely matches our detection,
while a $\nuddot$ contribution,
assuming a braking index $n\equiv\nus\nuddot/\nudot^2=3$,
would only change the $\nudot$ from \citet{halpernetal12} by 0.3~percent to
$\nudot\approx-6.317\times10^{-11}\mbox{ Hz s$^{-1}$}$.

\begin{figure}
\begin{center}
\includegraphics[width=\columnwidth]{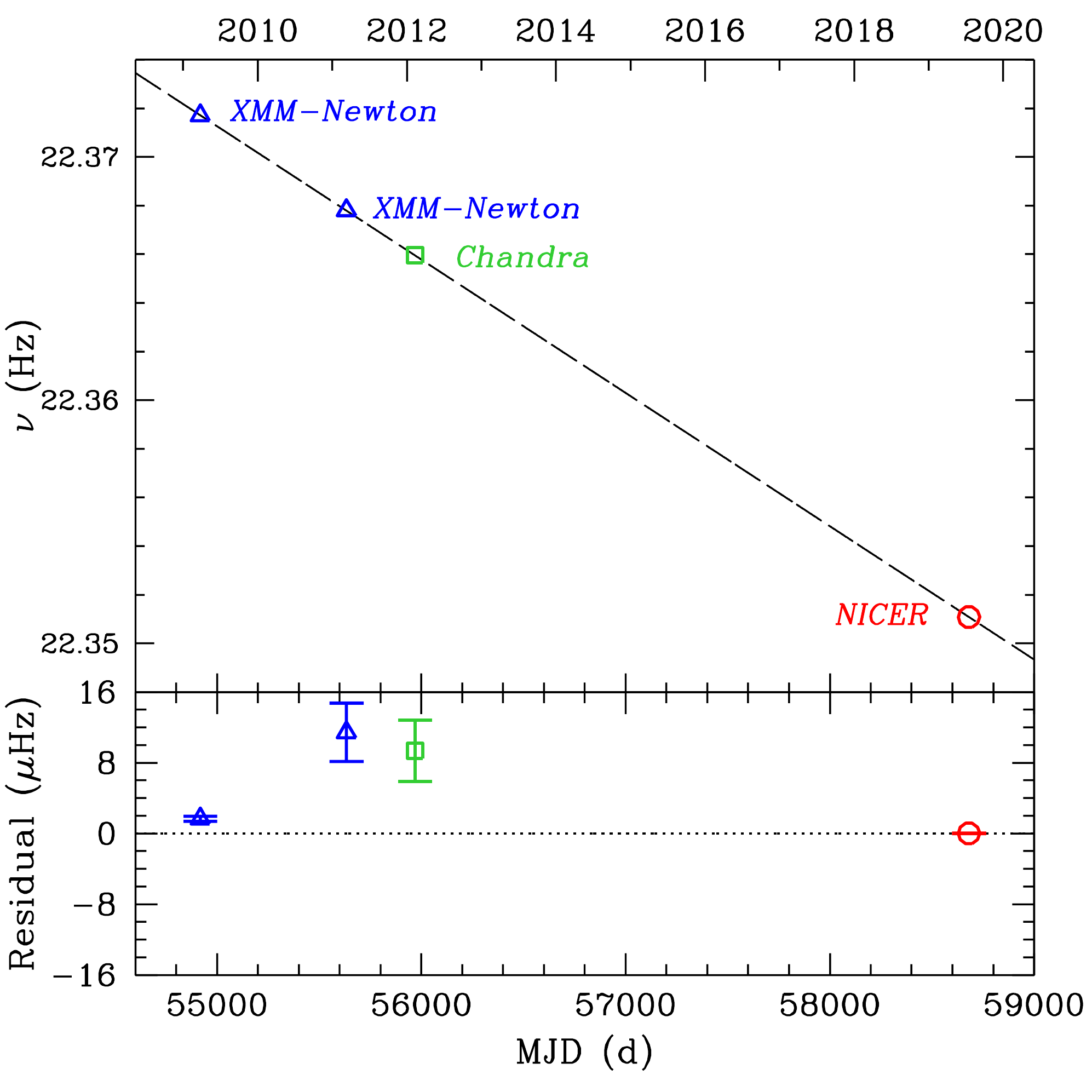}
\caption{Top: Spin frequency of \psr\ as measured using
\XMM\ in 2009 and 2011, \Chandra\ in 2012, and \NICER\ in 2019.
Dashed line shows a linear model fit to the spin frequencies,
with best-fit slope $\nudot=-6.3445\times 10^{-11}\mbox{ Hz s$^{-1}$}$.
Bottom: Spin frequency residual after subtracting off the best-fit
linear model.  Error bars are 1$\sigma$ uncertainty in measured $\nus$.
}
\label{fig:spin}
\end{center}
\end{figure}

We improve detection significance and refine the timing model
using a Markov Chain Monte Carlo (MCMC)
refiner in combination with the PINT
package\footnote{\url{https://github.com/nanograv/PINT}} and
the $H$-statistic as a measure of significance
\citep{dejageretal89,dejageretal10}.
Here we ignore the first observation (ObsID 1020440101) because
of its short exposure time and because it is too far removed in time
(ten months) from the other observations to smoothly connect with
the timing model obtained from these later observations.
We add an increasing number of terms in frequency derivative
and stop at $\nuddot$ (see below) since the addition of $\dddot{\nus}$
only yields a marginal $H$-test improvement.
We use TEMPO2 \citep{hobbsetal06} with the \texttt{photons}
plugin\footnote{\url{http://www.physics.mcgill.ca/~aarchiba/photons_plug.html}}
to assign pulse phases to each event.
Then using the \texttt{ni$\_$Htest$\_$sortgti.py} script from the
NICERsoft package,
we determine the optimum energy range to be 3.0--9.2~keV (124,656 events),
which yields a maximum significance of $10.1\sigma$;
we note that the choice of lowest energy to include has a large
impact on detection significance while the choice of highest
energy produces very similar significance levels.
The resulting timing model has $\nus=22.351086\pm0.000002\mbox{ Hz}$,
$\nudot=(-6.07\pm0.01)\times10^{-11}\mbox{ Hz s$^{-1}$}$, and
$\nuddot=(-1.0221\pm0.0001)\times10^{-17}\mbox{ Hz s$^{-2}$}$.
The large negative value of $\nuddot$ implies that, not only is
$\nudot$ becoming more negative, but $\nudot$ changes significantly
over the short timespan covered by the timing model, i.e.,
$|\Delta\nudot/\nudot|=0.55$ in 37.5~days.
Such a $\nuddot$ cannot be the pulsar's long-term value since it
would produce a different $\nudot$ from that determined about a
decade ago.
Figure~\ref{fig:spin} shows the spin frequency of \psr\ over the last
ten years,
i.e., those from 2009--2012 measured by \citet{halpernetal12} and our
2019 narrow windows search result (see below).
A simple linear model fit to these $\nus$ yields a best-fit spin-down
rate of $\nudot=(-6.3445\pm0.0004)\times 10^{-11}\mbox{ Hz s$^{-1}$}$,
which is within 3$\sigma$ of that determined by \citet{halpernetal12},
i.e., $\nudot=(-6.3335\pm0.0032)\times 10^{-11}\mbox{ Hz s$^{-1}$}$,
and is clearly the long-term spin-down rate of the pulsar.
In addition to the blind searches described above, we perform a
search in narrow windows around the expected $\nus$ and $\nudot$
(i.e., $\nus=[22.35108\mbox{ Hz},22.35108995\mbox{ Hz}]$ with steps
of $5\times 10^{-8}\mbox{ Hz}$ and
$\nudot=[-7\times10^{-11}\mbox{ Hz s$^{-1}$},-6\times10^{-11}\mbox{ Hz s$^{-1}$}]$
in steps of $1\times 10^{-13}\mbox{ Hz s$^{-1}$}$)
and assuming $\nuddot=0$.
Search results (with $H$-test value of 72) are presented in
Table~\ref{tab:timingmodel}, including a
$\nudot=(-6.428\pm0.003)\times 10^{-11}\mbox{ Hz s$^{-1}$}$,
which differs but is much closer to the long-term value.

\begin{table}
\centering
\caption{Timing parameters of \psr.
Two sets of parameters are provided:
The first is from a narrow windows search of only the 2019 \NICER\ data
(see text); the second is from a linear fit of $\nus$ from 2009--2012
and $\nus$ from the first set.
Number in parentheses is $1\sigma$ error in last digit.}
\label{tab:timingmodel}
\begin{tabular}{ll}
\hline
Parameter & Value \\
\hline
R.A. (J2000) & $18^{\rm h}13^{\rm m}35.\!\!^{\rm s}112$ \\
Decl. (J2000) & $-17^\circ49\arcmin57.\!\!\arcsec57$ \\
Position epoch (MJD) & 57544 \\
Timing reference epoch (MJD) & $58681.04405092593$ \\
Spin frequency $\nus$ (Hz) & 22.351083818(17) \\
\multicolumn{2}{c}{2019 June--2019 August} \\
Timespan of model (MJD) & 58662.3--58699.8 \\
Frequency derivative $\nudot$ (Hz s$^{-1}$) & $-6.4283(33)\times10^{-11}$ \\
\multicolumn{2}{c}{2009 March--2019 August} \\
Timespan of model (MJD) & 54918.14--58699.8 \\
Frequency derivative $\nudot$ (Hz s$^{-1}$) & $-6.34450(44)\times10^{-11}$ \\
\hline
\end{tabular}
\end{table}

To investigate possible changes in the timing parameters within
the \NICER\ dataset, we break the dataset into three segments, i.e.,
ObsIDs 2579030101--3 (2019 June 28--30) for 52~ks of total exposure,
ObsIDs 2579030201--4 (July 10--13) for 50~ks, and
ObsIDs 2579030301--6 (July 30--August 4) for 49~ks.
The first \NICER\ observation of 6~ks exposure time (ObsID 1020440101)
is dropped since it is not long enough to yield an independent
detection on its own.
A blind search is run on each segment independently, with the
same search parameters as given above.  A clear detection is made
only in the third segment, and the best pulsation candidate has
$\nus=22.35099411\mbox{ Hz}$ and
$\nudot=-6.217\times10^{-11}\mbox{ Hz s$^{-1}$}$, which are
consistent with results from the merged dataset search,
with a p-value of $5.62\times10^{-5}$.
Meanwhile, performing a search of each segment in a narrow window,
with steps in $\nus$ of $1\times 10^{-7}\mbox{ Hz}$ but using a
fixed $\nudot=-6.428\times10^{-11}\mbox{ Hz s$^{-1}$}$, yields a
detection in all three segments ($H$-test values $>30$).
There is also evidence from the structure of the significance peaks
in $\nu$ in the third segment that a glitch of magnitude
$\Delta\nus\approx3\mbox{ $\mu$Hz}$ occurred on MJD~58698,
but data limitations prevent a more definitive conclusion.

\begin{figure}
\begin{center}
\includegraphics[width=\columnwidth]{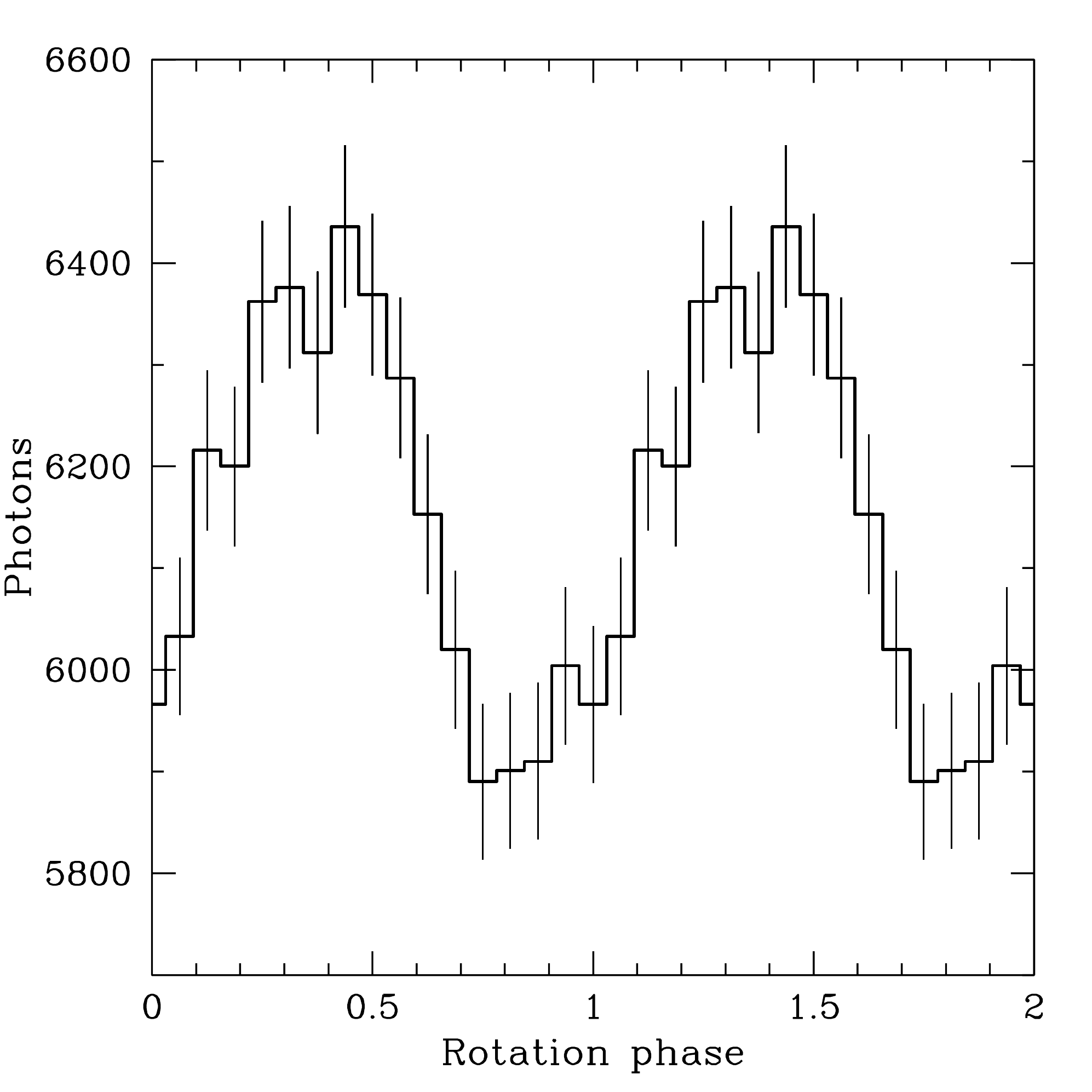}
\caption{Pulse profile (3--9.2~keV) of \psr\ using the 2019-only timing model
given in Table~\ref{tab:timingmodel}.
Two rotation cycles are shown, with 16 bins per cycle.
Error bars are $1\sigma$.}
\label{fig:pulseprofile}
\end{center}
\end{figure}

In summary, Table~\ref{tab:timingmodel} presents the final parameters
of two timing models we derived above.
The first is based on only the narrow windows search of the 37~d span
in 2019 of \NICER\ data.
The second is based on the linear fit to the long-term ten-year spin
frequency evolution starting from 2009
to our 2019 measurement using \NICER\ and shown in Figure~\ref{fig:spin}.
The main distinction between the two is
$\nudot=-6.428\times 10^{-11}\mbox{ Hz s$^{-1}$}$ in the former and
$\nudot=-6.3445\times 10^{-11}\mbox{ Hz s$^{-1}$}$ in the latter.
We also show in Figure~\ref{fig:pulseprofile} the 3--9.2~keV pulse
profile using \NICER\ data and the \NICER-only timing model;
because \NICER\ cannot spatially resolve PWN and supernova remnant
contributions to the unpulsed component of the pulse profile,
an estimate of the pulsed fraction would be unreliable.
There are several possible explanations for the differing spin-down
values, such as the pulsar's timing
behavior is affected by timing noise, glitches, and/or mode switching.
First, young pulsars, such as \psr, tend to exhibit timing noise
\citep{hobbsetal10,shannonetal10,espinozaetal17}.
For example, PSR~J1124$-$5916 in the supernova remnant G292.0+1.8
has comparable spin properties but is noisy; it also has a low
negative braking index and glitched at least once \citep{rayetal11}.
As for glitches,
\psr\ could glitch at a rate as high as once every $\approx 200\mbox{ d}$,
based on its spin-down rate $\nudot$ \citep{haskelletal12}.
In this case, there is a $\approx 15$~percent probability one glitch
occurred during the \NICER\ observations if glitch occurrence follows
Poisson statistics, and
we indeed have tentative evidence for the occurrence of such a glitch.
However, \citet{fuentesetal17,fuentesetal19} find that, while
glitch activity correlates with spin-down rate for
$|\nudot|<10^{-10.5}\mbox{ Hz s$^{-1}$}$ and is highest at the
top of this range, some pulsars with even greater $|\nudot|$ show
lower glitch activity.
Thus \psr\ could have a relatively low rate of glitches since
it has a $|\nudot|=10^{-10.2}\mbox{ Hz s$^{-1}$}$, but this is
far from certain.
Lastly, some pulsars switch between radio emission states, such
that their spin-down rate can be up to 50~percent higher when in an
active state \citep{krameretal06,lyneetal10,hermsenetal17}.
In the case of \psr, the best-fit long-term spin-down rate differs
from that of the timing model by only 1~percent.
\citet{dzibetal18} indicate pulsed radio emission may have been
detected but only at high frequencies due to extreme scattering.
An interesting comparison could potentially be made between \psr\
and the third highest $\Edot$ pulsar, PSR~B0540$-$69, which glitches
and has a similar age ($\sim 1000\mbox{ yr}$) and spin rate
($=19.7\mbox{ Hz}$) and $\nudot$ and $\nuddot$ that changed from
$-1.87\times 10^{-10}\mbox{ Hz s$^{-1}$}$ to
$-2.53\times 10^{-10}\mbox{ Hz s$^{-1}$}$ and
$3.7\times 10^{-21}\mbox{ Hz s$^{-2}$}$ to
$0.1\times 10^{-21}\mbox{ Hz s$^{-2}$}$, respectively,
due possibly to a switch between radio emission states
\citep{ferdmanetal15,marshalletal15,marshalletal16}.
More extensive X-ray monitoring and continued searches in radio
may be able to answer this issue.

\section{Discussions} \label{sec:discuss}

In this work, we analyze \Chandra\ and \NICER\ observations of
the highly energetic rotation-powered pulsar \psr.
The ten years between \Chandra\ observations allow us to measure a
pulsar proper motion $\mu_{\rm R.A.}=-0.\!\!\arcsec067\mbox{ yr$^{-1}$}$
and $\mu_{\rm decl.}=-0.\!\!\arcsec014\mbox{ yr$^{-1}$}$,
which implies a transverse velocity
$v_\perp=1600\mbox{ km s$^{-1}$}\,(d/\mbox{ 5 kpc})$;
note the distance is uncertain, with a range of 3--5~kpc
\citep{messineoetal11,halpernetal12}.
\Chandra\ spectra of the pulsar and its PWN can each be well-fit by
an absorbed power law model and reveal no evidence of significant
changes in flux and model parameters between observations.
However, we find that the X-ray absorption
($\NH=13.1\times 10^{22}\mbox{ cm$^{-2}$}$) is higher than that found
in early studies but in agreement with more recent works.
We detect pulsations at the 44.7~ms spin period of \psr\ using recent
\NICER\ data.
We find that the spin-down rate determined over
the past decade differs from that measured over a
month with the timing model.

The spin period of \psr\ has only been previously measured using \XMM\
in 2009 and 2011 and \Chandra\ in 2012 and not found in radio or
using \Fermi\ \citep{halpernetal12,kuiperhermsen15}.
The pulsar is a target of interest for gravitational wave (GW) searches.
These searches are beginning to achieve meaningful constraints on
possible energy loss due to GW emission in the case of \psr\
\citep{abbottetal17,abbottetal19b}.
The most sensitive searches that can be conducted are those that have
contemporaneous electromagnetic timing models \citep{abbottetal19a}.
If such a timing model had been available for \psr, an improvement
of $\sim$5 on upper limits to GW strain and $\sim$25 to GW energy
would have been possible over those obtained in the less sensitive
search by \citet{abbottetal19b}.
Efforts are underway to search the latest, most sensitive GW data
from the third observing run (O3), which collected data from
2019 April 1 to 2020 March 27.
The contemporaneous timing model provided here using \NICER\ data will
enable improved limits on GW emission from this highly energetic pulsar.

\section*{Acknowledgements}

The authors thank the anonymous referee for comments which led to
improvements in the manuscript.
WCGH thanks D.L. Kaplan and G.G. Pavlov for helpful comments on
astrometry analysis, P.S. Ray for advice on timing analysis, and
W.A. Majid for comments on an early draft.
WCGH appreciates use of computer facilities at the Kavli Institute for
Particle Astrophysics and Cosmology.
WCGH acknowledges support through grant 80NSSC19K1444 from NASA.
SG acknowledges support of CNES.
MB is partially supported by Polish NCN grant no.~2017/26/M/ST9/00978.
CME acknowledges funding from ANID grant FONDECYT/Regular 1171421.
CM is supported by an appointment to the NASA Postdoctoral Program at
Marshall Space Flight Center, which is administered by Universities
Space Research Association under contract with NASA.


\section*{Data availability}

The data underlying this article will be shared on reasonable
request to the corresponding author.


\bibliographystyle{mnras}
\bibliography{psrj1813}

\begin{thebibliography}{}
\makeatletter
\relax
\def\mn@urlcharsother{\let\do\@makeother \do\$\do\&\do\#\do\^\do\_\do\%\do\~}
\def\mn@doi{\begingroup\mn@urlcharsother \@ifnextchar [ {\mn@doi@}
  {\mn@doi@[]}}
\def\mn@doi@[#1]#2{\def\@tempa{#1}\ifx\@tempa\@empty \href
  {http://dx.doi.org/#2} {doi:#2}\else \href {http://dx.doi.org/#2} {#1}\fi
  \endgroup}
\def\mn@eprint#1#2{\mn@eprint@#1:#2::\@nil}
\def\mn@eprint@arXiv#1{\href {http://arxiv.org/abs/#1} {{\tt arXiv:#1}}}
\def\mn@eprint@dblp#1{\href {http://dblp.uni-trier.de/rec/bibtex/#1.xml}
  {dblp:#1}}
\def\mn@eprint@#1:#2:#3:#4\@nil{\def\@tempa {#1}\def\@tempb {#2}\def\@tempc
  {#3}\ifx \@tempc \@empty \let \@tempc \@tempb \let \@tempb \@tempa \fi \ifx
  \@tempb \@empty \def\@tempb {arXiv}\fi \@ifundefined
  {mn@eprint@\@tempb}{\@tempb:\@tempc}{\expandafter \expandafter \csname
  mn@eprint@\@tempb\endcsname \expandafter{\@tempc}}}

\bibitem[\protect\citeauthoryear{{{Abbott} et al.}}{{{Abbott} et
  al.}}{2017}]{abbottetal17}
{{Abbott} et al.} 2017, \mn@doi [\prd] {10.1103/PhysRevD.96.122006}, \href
  {https://ui.adsabs.harvard.edu/abs/2017PhRvD..96l2006A} {96, 122006}

\bibitem[\protect\citeauthoryear{{{Abbott} et al.}}{{{Abbott} et
  al.}}{2019a}]{abbottetal19b}
{{Abbott} et al.} 2019a, \mn@doi [\prd] {10.1103/PhysRevD.99.122002}, \href
  {https://ui.adsabs.harvard.edu/abs/2019PhRvD..99l2002A} {99, 122002}

\bibitem[\protect\citeauthoryear{{{Abbott} et al.}}{{{Abbott} et
  al.}}{2019b}]{abbottetal19a}
{{Abbott} et al.} 2019b, \mn@doi [\apj] {10.3847/1538-4357/ab20cb}, \href
  {https://ui.adsabs.harvard.edu/abs/2019ApJ...879...10A} {879, 10}

\bibitem[\protect\citeauthoryear{{Abdo} et~al.,}{{Abdo}
  et~al.}{2009}]{abdoetal09}
{Abdo} A.~A.,  et~al., 2009, \mn@doi [Science] {10.1126/science.1175558}, \href
  {https://ui.adsabs.harvard.edu/abs/2009Sci...325..840A} {325, 840}

\bibitem[\protect\citeauthoryear{{Aharonian} et~al.,}{{Aharonian}
  et~al.}{2005}]{aharonianetal05}
{Aharonian} F.,  et~al., 2005, \mn@doi [Science] {10.1126/science.1108643},
  \href {https://ui.adsabs.harvard.edu/abs/2005Sci...307.1938A} {307, 1938}

\bibitem[\protect\citeauthoryear{{Arnaud}}{{Arnaud}}{1996}]{arnaud96}
{Arnaud} K.~A.,  1996, in {Jacoby} G.~H.,  {Barnes} J.,  eds,  Astronomical
  Society of the Pacific Conference Series Vol. 101, Astronomical Data Analysis
  Software and Systems V. p.~17

\bibitem[\protect\citeauthoryear{{Atwood}, {Ziegler}, {Johnson}  \&
  {Baughman}}{{Atwood} et~al.}{2006}]{atwoodetal06}
{Atwood} W.~B.,  {Ziegler} M.,  {Johnson} R.~P.,   {Baughman} B.~M.,  2006,
  \mn@doi [\apjl] {10.1086/510018}, \href
  {https://ui.adsabs.harvard.edu/abs/2006ApJ...652L..49A} {652, L49}

\bibitem[\protect\citeauthoryear{{Becker}}{{Becker}}{2009}]{becker09}
{Becker} W.,  2009, {X-Ray Emission from Pulsars and Neutron Stars}.
p.~91, \mn@doi{10.1007/978-3-540-76965-1_6}

\bibitem[\protect\citeauthoryear{{Becker} \& {Tr\"umper}}{{Becker} \&
  {Tr\"umper}}{1997}]{beckertrumper97}
{Becker} W.,  {Tr\"umper} J.,  1997, \aap, \href
  {https://ui.adsabs.harvard.edu/abs/1997A&A...326..682B} {326, 682}

\bibitem[\protect\citeauthoryear{{Brogan}, {Gaensler}, {Gelfand}, {Lazendic},
  {Lazio}, {Kassim}  \& {McClure-Griffiths}}{{Brogan}
  et~al.}{2005}]{broganetal05}
{Brogan} C.~L.,  {Gaensler} B.~M.,  {Gelfand} J.~D.,  {Lazendic} J.~S.,
  {Lazio} T.~J.~W.,  {Kassim} N.~E.,   {McClure-Griffiths} N.~M.,  2005,
  \mn@doi [\apjl] {10.1086/491471}, \href
  {https://ui.adsabs.harvard.edu/abs/2005ApJ...629L.105B} {629, L105}

\bibitem[\protect\citeauthoryear{{Dang} et~al.,}{{Dang}
  et~al.}{2020}]{dangetal20}
{Dang} S.~J.,  et~al., 2020, \mn@doi [\apj] {10.3847/1538-4357/ab9082}, \href
  {https://ui.adsabs.harvard.edu/abs/2020arXiv200502200D} {896, 17}

\bibitem[\protect\citeauthoryear{{Deller} et~al.,}{{Deller}
  et~al.}{2019}]{delleretal19}
{Deller} A.~T.,  et~al., 2019, \mn@doi [\apj] {10.3847/1538-4357/ab11c7}, \href
  {https://ui.adsabs.harvard.edu/abs/2019ApJ...875..100D} {875, 100}

\bibitem[\protect\citeauthoryear{{Dzib}, {Loinard}  \& {Rodr{\'\i}guez}}{{Dzib}
  et~al.}{2010}]{dzibetal10}
{Dzib} S.,  {Loinard} L.,   {Rodr{\'\i}guez} L.~F.,  2010, \rmxaa, \href
  {https://ui.adsabs.harvard.edu/abs/2010RMxAA..46..153D} {46, 153}

\bibitem[\protect\citeauthoryear{{Dzib}, {Rodr{\'\i}guez}, {Karuppusamy},
  {Loinard}  \& {Medina}}{{Dzib} et~al.}{2018}]{dzibetal18}
{Dzib} S.~A.,  {Rodr{\'\i}guez} L.~F.,  {Karuppusamy} R.,  {Loinard} L.,
  {Medina} S.-N.~X.,  2018, \mn@doi [\apj] {10.3847/1538-4357/aada07}, \href
  {https://ui.adsabs.harvard.edu/abs/2018ApJ...866..100D} {866, 100}

\bibitem[\protect\citeauthoryear{{Enoto}, {Kisaka}  \& {Shibata}}{{Enoto}
  et~al.}{2019}]{enotoetal19}
{Enoto} T.,  {Kisaka} S.,   {Shibata} S.,  2019, \mn@doi [Reports on Progress
  in Physics] {10.1088/1361-6633/ab3def}, \href
  {https://ui.adsabs.harvard.edu/abs/2019RPPh...82j6901E} {82, 106901}

\bibitem[\protect\citeauthoryear{{Espinoza}, {Lyne}  \& {Stappers}}{{Espinoza}
  et~al.}{2017}]{espinozaetal17}
{Espinoza} C.~M.,  {Lyne} A.~G.,   {Stappers} B.~W.,  2017, \mn@doi [\mnras]
  {10.1093/mnras/stw3081}, \href
  {https://ui.adsabs.harvard.edu/abs/2017MNRAS.466..147E} {466, 147}

\bibitem[\protect\citeauthoryear{{Ferdman}, {Archibald}  \& {Kaspi}}{{Ferdman}
  et~al.}{2015}]{ferdmanetal15}
{Ferdman} R.~D.,  {Archibald} R.~F.,   {Kaspi} V.~M.,  2015, \mn@doi [\apj]
  {10.1088/0004-637X/812/2/95}, \href
  {https://ui.adsabs.harvard.edu/abs/2015ApJ...812...95F} {812, 95}

\bibitem[\protect\citeauthoryear{{Fruscione} et~al.,}{{Fruscione}
  et~al.}{2006}]{fruscioneetal06}
{Fruscione} A.,  et~al., 2006, in Society of Photo-Optical Instrumentation
  Engineers (SPIE) Conference Series. p. 62701V, \mn@doi{10.1117/12.671760}

\bibitem[\protect\citeauthoryear{{Fuentes}, {Espinoza}, {Reisenegger}, {Shaw},
  {Stappers}  \& {Lyne}}{{Fuentes} et~al.}{2017}]{fuentesetal17}
{Fuentes} J.~R.,  {Espinoza} C.~M.,  {Reisenegger} A.,  {Shaw} B.,  {Stappers}
  B.~W.,   {Lyne} A.~G.,  2017, \mn@doi [\aap] {10.1051/0004-6361/201731519},
  \href {https://ui.adsabs.harvard.edu/abs/2017A&A...608A.131F} {608, A131}

\bibitem[\protect\citeauthoryear{{Fuentes}, {Espinoza}  \&
  {Reisenegger}}{{Fuentes} et~al.}{2019}]{fuentesetal19}
{Fuentes} J.~R.,  {Espinoza} C.~M.,   {Reisenegger} A.,  2019, \mn@doi [\aap]
  {10.1051/0004-6361/201935939}, \href
  {https://ui.adsabs.harvard.edu/abs/2019A&A...630A.115F} {630, A115}

\bibitem[\protect\citeauthoryear{{Funk} et~al.,}{{Funk}
  et~al.}{2007}]{funketal07}
{Funk} S.,  et~al., 2007, \mn@doi [\aap] {10.1051/0004-6361:20066779}, \href
  {https://ui.adsabs.harvard.edu/abs/2007A&A...470..249F} {470, 249}

\bibitem[\protect\citeauthoryear{{Gotthelf} \& {Halpern}}{{Gotthelf} \&
  {Halpern}}{2009}]{gotthelfhalpern09}
{Gotthelf} E.~V.,  {Halpern} J.~P.,  2009, \mn@doi [\apjl]
  {10.1088/0004-637X/700/2/L158}, \href
  {https://ui.adsabs.harvard.edu/abs/2009ApJ...700L.158G} {700, L158}

\bibitem[\protect\citeauthoryear{{HI4PI Collaboration} et~al.,}{{HI4PI
  Collaboration} et~al.}{2016}]{hi4pi16}
{HI4PI Collaboration} et~al., 2016, \mn@doi [\aap]
  {10.1051/0004-6361/201629178}, \href
  {https://ui.adsabs.harvard.edu/abs/2016A&A...594A.116H} {594, A116}

\bibitem[\protect\citeauthoryear{{Halpern}, {Gotthelf}  \& {Camilo}}{{Halpern}
  et~al.}{2012}]{halpernetal12}
{Halpern} J.~P.,  {Gotthelf} E.~V.,   {Camilo} F.,  2012, \mn@doi [\apjl]
  {10.1088/2041-8205/753/1/L14}, \href
  {https://ui.adsabs.harvard.edu/abs/2012ApJ...753L..14H} {753, L14}

\bibitem[\protect\citeauthoryear{{Haskell}, {Pizzochero}  \&
  {Sidery}}{{Haskell} et~al.}{2012}]{haskelletal12}
{Haskell} B.,  {Pizzochero} P.~M.,   {Sidery} T.,  2012, \mn@doi [\mnras]
  {10.1111/j.1365-2966.2011.20080.x}, \href
  {https://ui.adsabs.harvard.edu/abs/2012MNRAS.420..658H} {420, 658}

\bibitem[\protect\citeauthoryear{{Helfand}, {Gotthelf}, {Halpern}, {Camilo},
  {Semler}, {Becker}  \& {White}}{{Helfand} et~al.}{2007}]{helfandetal07}
{Helfand} D.~J.,  {Gotthelf} E.~V.,  {Halpern} J.~P.,  {Camilo} F.,  {Semler}
  D.~R.,  {Becker} R.~H.,   {White} R.~L.,  2007, \mn@doi [\apj]
  {10.1086/519734}, \href
  {https://ui.adsabs.harvard.edu/abs/2007ApJ...665.1297H} {665, 1297}

\bibitem[\protect\citeauthoryear{{Hermsen} et~al.,}{{Hermsen}
  et~al.}{2017}]{hermsenetal17}
{Hermsen} W.,  et~al., 2017, \mn@doi [\mnras] {10.1093/mnras/stw3135}, \href
  {https://ui.adsabs.harvard.edu/abs/2017MNRAS.466.1688H} {466, 1688}

\bibitem[\protect\citeauthoryear{{Hobbs}, {Edwards}  \& {Manchester}}{{Hobbs}
  et~al.}{2006}]{hobbsetal06}
{Hobbs} G.~B.,  {Edwards} R.~T.,   {Manchester} R.~N.,  2006, \mn@doi [\mnras]
  {10.1111/j.1365-2966.2006.10302.x}, \href
  {https://ui.adsabs.harvard.edu/abs/2006MNRAS.369..655H} {369, 655}

\bibitem[\protect\citeauthoryear{{Hobbs}, {Lyne}  \& {Kramer}}{{Hobbs}
  et~al.}{2010}]{hobbsetal10}
{Hobbs} G.,  {Lyne} A.~G.,   {Kramer} M.,  2010, \mn@doi [\mnras]
  {10.1111/j.1365-2966.2009.15938.x}, \href
  {https://ui.adsabs.harvard.edu/abs/2010MNRAS.402.1027H} {402, 1027}

\bibitem[\protect\citeauthoryear{{Kargaltsev}, {Pavlov}, {Klingler}  \&
  {Rangelov}}{{Kargaltsev} et~al.}{2017}]{kargaltsevetal17}
{Kargaltsev} O.,  {Pavlov} G.~G.,  {Klingler} N.,   {Rangelov} B.,  2017,
  \mn@doi [Journal of Plasma Physics] {10.1017/S0022377817000630}, \href
  {https://ui.adsabs.harvard.edu/abs/2017JPlPh..83e6301K} {83, 635830501}

\bibitem[\protect\citeauthoryear{{Kramer}, {Lyne}, {O'Brien}, {Jordan}  \&
  {Lorimer}}{{Kramer} et~al.}{2006}]{krameretal06}
{Kramer} M.,  {Lyne} A.~G.,  {O'Brien} J.~T.,  {Jordan} C.~A.,   {Lorimer}
  D.~R.,  2006, \mn@doi [Science] {10.1126/science.1124060}, \href
  {https://ui.adsabs.harvard.edu/abs/2006Sci...312..549K} {312, 549}

\bibitem[\protect\citeauthoryear{{Kuiper} \& {Hermsen}}{{Kuiper} \&
  {Hermsen}}{2015}]{kuiperhermsen15}
{Kuiper} L.,  {Hermsen} W.,  2015, \mn@doi [\mnras] {10.1093/mnras/stv426},
  \href {https://ui.adsabs.harvard.edu/abs/2015MNRAS.449.3827K} {449, 3827}

\bibitem[\protect\citeauthoryear{{Lyne}, {Hobbs}, {Kramer}, {Stairs}  \&
  {Stappers}}{{Lyne} et~al.}{2010}]{lyneetal10}
{Lyne} A.,  {Hobbs} G.,  {Kramer} M.,  {Stairs} I.,   {Stappers} B.,  2010,
  \mn@doi [Science] {10.1126/science.1186683}, \href
  {https://ui.adsabs.harvard.edu/abs/2010Sci...329..408L} {329, 408}

\bibitem[\protect\citeauthoryear{{Manchester}, {Hobbs}, {Teoh}  \&
  {Hobbs}}{{Manchester} et~al.}{2005}]{manchesteretal05}
{Manchester} R.~N.,  {Hobbs} G.~B.,  {Teoh} A.,   {Hobbs} M.,  2005, \mn@doi
  [\aj] {10.1086/428488}, \href
  {https://ui.adsabs.harvard.edu/abs/2005AJ....129.1993M} {129, 1993}

\bibitem[\protect\citeauthoryear{{Marelli}, {De Luca}  \& {Caraveo}}{{Marelli}
  et~al.}{2011}]{marellietal11}
{Marelli} M.,  {De Luca} A.,   {Caraveo} P.~A.,  2011, \mn@doi [\apj]
  {10.1088/0004-637X/733/2/82}, \href
  {https://ui.adsabs.harvard.edu/abs/2011ApJ...733...82M} {733, 82}

\bibitem[\protect\citeauthoryear{{Marshall}, {Guillemot}, {Harding}, {Martin}
  \& {Smith}}{{Marshall} et~al.}{2015}]{marshalletal15}
{Marshall} F.~E.,  {Guillemot} L.,  {Harding} A.~K.,  {Martin} P.,   {Smith}
  D.~A.,  2015, \mn@doi [\apjl] {10.1088/2041-8205/807/2/L27}, \href
  {https://ui.adsabs.harvard.edu/abs/2015ApJ...807L..27M} {807, L27}

\bibitem[\protect\citeauthoryear{{Marshall}, {Guillemot}, {Harding}, {Martin}
  \& {Smith}}{{Marshall} et~al.}{2016}]{marshalletal16}
{Marshall} F.~E.,  {Guillemot} L.,  {Harding} A.~K.,  {Martin} P.,   {Smith}
  D.~A.,  2016, \mn@doi [\apjl] {10.3847/2041-8205/827/2/L39}, \href
  {https://ui.adsabs.harvard.edu/abs/2016ApJ...827L..39M} {827, L39}

\bibitem[\protect\citeauthoryear{{Messineo}, {Davies}, {Figer}, {Kudritzki},
  {Valenti}, {Trombley}, {Najarro}  \& {Rich}}{{Messineo}
  et~al.}{2011}]{messineoetal11}
{Messineo} M.,  {Davies} B.,  {Figer} D.~F.,  {Kudritzki} R.~P.,  {Valenti} E.,
   {Trombley} C.,  {Najarro} F.,   {Rich} R.~M.,  2011, \mn@doi [\apj]
  {10.1088/0004-637X/733/1/41}, \href
  {https://ui.adsabs.harvard.edu/abs/2011ApJ...733...41M} {733, 41}

\bibitem[\protect\citeauthoryear{{Ray} et~al.,}{{Ray} et~al.}{2011}]{rayetal11}
{Ray} P.~S.,  et~al., 2011, \mn@doi [\apjs] {10.1088/0067-0049/194/2/17}, \href
  {https://ui.adsabs.harvard.edu/abs/2011ApJS..194...17R} {194, 17}

\bibitem[\protect\citeauthoryear{{Saz Parkinson} et~al.,}{{Saz Parkinson}
  et~al.}{2010}]{sazparkinsonetal10}
{Saz Parkinson} P.~M.,  et~al., 2010, \mn@doi [\apj]
  {10.1088/0004-637X/725/1/571}, \href
  {https://ui.adsabs.harvard.edu/abs/2010ApJ...725..571S} {725, 571}

\bibitem[\protect\citeauthoryear{{Shannon} \& {Cordes}}{{Shannon} \&
  {Cordes}}{2010}]{shannonetal10}
{Shannon} R.~M.,  {Cordes} J.~M.,  2010, \mn@doi [\apj]
  {10.1088/0004-637X/725/2/1607}, \href
  {https://ui.adsabs.harvard.edu/abs/2010ApJ...725.1607S} {725, 1607}

\bibitem[\protect\citeauthoryear{{Townsley}, {Broos}, {Garmire}, {Anderson},
  {Feigelson}, {Naylor}  \& {Povich}}{{Townsley} et~al.}{2018}]{townsleyetal18}
{Townsley} L.~K.,  {Broos} P.~S.,  {Garmire} G.~P.,  {Anderson} G.~E.,
  {Feigelson} E.~D.,  {Naylor} T.,   {Povich} M.~S.,  2018, \mn@doi [\apjs]
  {10.3847/1538-4365/aaaf67}, \href
  {https://ui.adsabs.harvard.edu/abs/2018ApJS..235...43T} {235, 43}

\bibitem[\protect\citeauthoryear{{Ubertini} et~al.,}{{Ubertini}
  et~al.}{2005}]{ubertinietal05}
{Ubertini} P.,  et~al., 2005, \mn@doi [\apjl] {10.1086/447766}, \href
  {https://ui.adsabs.harvard.edu/abs/2005ApJ...629L.109U} {629, L109}

\bibitem[\protect\citeauthoryear{{Verner}, {Ferland}, {Korista}  \&
  {Yakovlev}}{{Verner} et~al.}{1996}]{verneretal96}
{Verner} D.~A.,  {Ferland} G.~J.,  {Korista} K.~T.,   {Yakovlev} D.~G.,  1996,
  \mn@doi [\apj] {10.1086/177435}, \href
  {https://ui.adsabs.harvard.edu/abs/1996ApJ...465..487V} {465, 487}

\bibitem[\protect\citeauthoryear{{Wilms}, {Allen}  \& {McCray}}{{Wilms}
  et~al.}{2000}]{wilmsetal00}
{Wilms} J.,  {Allen} A.,   {McCray} R.,  2000, \mn@doi [\apj] {10.1086/317016},
  \href {http://adsabs.harvard.edu/abs/2000ApJ...542..914W} {542, 914}

\bibitem[\protect\citeauthoryear{{de Jager} \& {B{\"u}sching}}{{de Jager} \&
  {B{\"u}sching}}{2010}]{dejageretal10}
{de Jager} O.~C.,  {B{\"u}sching} I.,  2010, \mn@doi [\aap]
  {10.1051/0004-6361/201014362}, \href
  {https://ui.adsabs.harvard.edu/abs/2010A&A...517L...9D} {517, L9}

\bibitem[\protect\citeauthoryear{{de Jager}, {Raubenheimer}  \&
  {Swanepoel}}{{de Jager} et~al.}{1989}]{dejageretal89}
{de Jager} O.~C.,  {Raubenheimer} B.~C.,   {Swanepoel} J.~W.~H.,  1989, \aap,
  \href {https://ui.adsabs.harvard.edu/abs/1989A&A...221..180D} {221, 180}

\makeatother
\end{thebibliography}


\bsp	
\label{lastpage}
\end{document}